\documentclass[10pt,journal]{IEEEtran}
\usepackage[gen]{eurosym}
\usepackage{multirow}
\usepackage{graphicx}
\usepackage{xcolor}
\usepackage{enumitem}
\usepackage{url}
\makeatletter
\g@addto@macro{\UrlBreaks}{\UrlOrds}
\makeatother
\ifCLASSINFOpdf
\begin{document}

\title{RSSI Fingerprinting-based Localization Using Machine Learning in LoRa Networks}
\author{Mahnoor~Anjum,
	Muhammad~Abdullah~Khan,
        Syed~Ali~Hassan,
        Aamir~Mahmood,
        Hassaan~Khaliq~Qureshi,
        and~Mikael~Gidlund
\thanks{M.~Anjum, M. A. Khan, S. A. Hassan, and H. K. Qureshi are with NUST, Pakistan.

A.~Mahmood, and M.~Gidlund are with Mid Sweden University, Sweden, e-mail: aamir.mahmood@miun.se.
}
\vspace{-20pt}}
\IEEEtitleabstractindextext{%
\begin{abstract}

The scale of wireless technologies penetration in our daily lives, primarily triggered by the Internet-of-things (IoT)-based smart cities, is beaconing the possibilities of novel localization and tracking techniques. Recently, low-power wide-area network (LPWAN) technologies have emerged as a solution to offer scalable wireless connectivity for smart city applications. LoRa is one such technology that provides energy efficiency and wide-area coverage.  {This article explores the use of intelligent machine learning techniques, such as support vector machines, spline models, decision trees, and ensemble learning, for received signal strength indicator (RSSI)-based ranging in LoRa networks, on a training dataset collected in two different environments: indoors and outdoors}. The suitable ranging model is then used to experimentally evaluate the accuracy of localization and tracking using trilateration in the studied environments.  {Later, we present the accuracy of LoRa-based positioning system (LPS) and compare it with the existing ZigBee, WiFi, and Bluetooth-based solutions.} In the end, we discuss the challenges of satellite-independent tracking systems and propose future directions to improve accuracy and provide deployment feasibility.
\end{abstract}

}

\maketitle

\IEEEdisplaynontitleabstractindextext

%
\IEEEpeerreviewmaketitle

\section{Introduction}
\IEEEPARstart{W}{ith} the conceptualization of smart cities, the Internet of-Things (IoT) market has seen a compound annual growth rate  of 14.2\% during 2011-2016 \cite{c1}{, while the IoT industry is expected to add \$11 trillion to the global economy by 2025 \cite{c3}. By every passing second, 127 new devices are getting connected to the Internet, and it is expected that there would be 125 billion IoT connected devices in 2030 \cite{c2}.} Smart cities refer to urban areas comprising several interconnected devices transferring sensor data to trigger actuators for managing, automating, and operating industrial, personal, and interpersonal operations efficiently and effectively. IoT connectivity solutions are built on diverse communication technologies, depending upon the data rate, range, power, and topology required by the application. The  {long-range} IoT solutions have traditionally used cellular technologies (2G/3G/4G) and the short-range solutions are engineered using Wi-Fi, Bluetooth Low Energy (BLE), ZigBee and 6LoWPAN technologies. 

Smart cities intensify the need for location-based services (LBS). LBS have gained popularity in the research community due to  {the implementation of selective availability  of the global positioning system (GPS) in the early 2000s. Selective availability was an international degradation of GPS for security reasons.} Satellite navigation (satnav)-based positioning systems have been used since the 1980s, and are currently the most popular geolocation service providers. However, they require a good view of the sky to develop a line-of-sight (LOS) fix with the satellites. Thus, satnav proves inefficient for dense-urban areas and indoor IoT applications such as asset tracking systems. 

Tracking-based IoT applications require low-power and long-range communication. Short-range IoT technologies such as BLE, ZigBee, and Wi-Fi work well in non-line-of-sight (NLOS) environments, but are limited in their area of coverage. Cellular technologies, though long-range in nature, are unsuitable for tracking purposes due to their energy inefficiency.  {These aspects together with the inherent limitations of GPS create a gap for LBS  in smart-city applications. In this respect, the introduction of low power wide area networks (LPWANs) for IoT enabling communications  is a promising opportunity to bridge the gap. LPWANs offer a trade-off between power consumption and the data rate; hence, are ideally suited for event-based IoT applications. Their suitability for LBS is further cemented by their wide-area of coverage; providing excellent infrastructure for smart-city positioning and tracking systems.} 

 {The main motivation behind this article is to explore the feasibility of LoRa networks, a type of LPWAN, to develop positioning system in indoor and outdoor environments. Towards this objective, first, we perform traditional regression-based and modern machine learning-based ranging of received signal strength (RSS) fingerprinting dataset in a multi-gateway LoRa network. Second, the best ranging models are used to experimentally evaluate trilateration-based positioning of static assets in the studied indoor and outdoor environments. We then discuss the results of the LoRa-based positioning system (LPS) in the light of the currently available systems and previous research works. Towards the end of this article, we discuss the potentials and deployment feasibility of an LPS solution and outline the open research directions and challenges.}

\section{Positioning Basis}
The positioning techniques compute the location of a device using a two-step process: a) range/angle estimation of a device from the exchanged radio frequency (RF) signals with the reference points (i.e., the base stations (BS) or gateways) located at known coordinates, and b) computing geometrical relations from these range measurements and coordinates of the reference points. In this section, we describe the available options within each step to build a feasible basis for our LoRa-based positioning system, which we term as LPS in the rest of this article.
\subsection{Ranging Techniques}
Ranging refers to the process of computing the distance estimate between a transmitter and a receiver using an RF signal. Depending on the measured property of the RF signal, the ranging  can be performed using one of the following techniques:
\begin{itemize}[leftmargin=*]
\item \textbf{Signal characteristics---RSSI}: Ranging based on a signal characteristic, often referred to as \textit{fingerprinting}, uses reference data to attribute distance to specific value(s) of signal characteristics. The most common characteristic is the received signal strength indicator (RSSI), which captures the power-distance relationship of a signal propagation environment. RSSI fingerprinting is extensively studied for Sigfox~\cite{c5}, Bluetooth~\cite{c10},  Wi-Fi~\cite{c6}, and ZigBee~\cite{c7} technologies.

\item \textbf{Time characteristics---ToA, TDoA}:  {Ranging is performed by measuring the time characteristics of the signal. Two commonly used techniques are time-of-arrival (ToA) and time-difference-of-arrival (TDoA) \cite{c8}. Assuming perfect synchronization between the device and the reference point, ToA scheme is based on knowing the exact time at which the signal is sent by the device and the exact time the signal arrives at the reference point, and the speed of the signal. Meanwhile, a two-way ToA-based ranging eliminates the need of synchronization. On the contrary, TDoA  only requires the time at which the signal is received at two `synchronized' reference points. The difference in arrival time is used to calculate the difference in distances between the device and the reference points.} 

\item \textbf{Angle measurements---AoA}: The technique uses the angle of arrival (AoA) measurements of RF signal from the device based on the spatial geometry of antenna array elements at reference points.   
\end{itemize}
It is worth noting that all the ranging techniques discussed above are highly affected by the propagation artifacts, i.e., multipath fading and NLOS conditions. Also, the techniques requiring time synchronization (i.e., TDoA, ToA) or antenna arrays (AoA) might not always be feasible for LPWAN systems.

\subsection{Positioning Techniques}
Computing the coordinates of a point (device) from the range/angle measurements is called as positioning. The device, seeking positioning, based on the distance or angle estimate of the reference points and their coordinates can solve for its unknown position through geometrical techniques (properties of triangles) as: 
\begin{enumerate}
    \item Trilateration or multilateration---using range estimate from signal characteristics (RSSI) or time characteristics (ToA, TDoA).
    \item Triangulation---using angle characteristics as AoA.
\end{enumerate}
The positioning can also be performed at a central system by executing these positioning algorithms on the range/angle estimates of the device collected from the reference points.   Note that the trilateration requires three reference points to find the position of a device in 2D space. Trilateration based positioning systems has been studied for ZigBee~\cite{c9}, Bluetooth low energy (BLE)~\cite{c10}, and other technologies.

\section{LPWANs for Positioning: Enabling Features in LoRaWAN}
The design of LPWANs is centered around simplified connectivity for massive IoT traffic, predominantly uplink and delay-tolerant, together with unprecedented connection density, coverage, and energy efficiency. The IMT-2020 vision for next-generation wireless networks identifies massive IoT connection density of $10^6/\textrm{km}^2$ with a device lifetime of more than 10 years~\cite{ITU_R}. This vision has stimulated the development race of LPWAN technologies in sub-1 GHz licensed and  unlicensed bands, since sub-1 GHz bands offer immunity against fading and attenuation, which leads to cover a wide-area with fewer BSs. As a result, NB-IoT emerged as a licensed band solution from 3GPP, while various new proprietary technologies (e.g., LoRa and Sigfox) surfaced in unlicensed bands~\cite{SIGFOX}. LoRa and Sigfox are witnessing adoption on exceptional scale for: 
\begin{enumerate}[label=(\alph*)]
    \item unique narrowband signaling; LoRa uses chirp spread spectrum (CSS) and Sigfox uses frequency hopping and ultra-narrowband signals. Narrowband signaling increases the number of available channels (although low data rates) as well as higher SNR (enhanced coverage) because of lower noise bandwidth,
    \item ALOHA-like random access, with restrictions primarily on the duty cycle, with minimal control overhead, and relaxed synchronization requirements, which among others reduces the energy consumption drastically,
    \item simplified backend network architecture, reducing the cost to own and manage a network.
\end{enumerate}
The increasing usage of LPWANs technologies is also increasing their popularity of network-based geolocation services, which are rapidly becoming a cost-effective use-case for LPWANs.

\subsection{LoRa and LoRAWAN}
For its wide-scale adoption in real-world networks, in the following, we briefly describe the components of a LoRaWAN network, in order to understand its key enabling features in general and for localization in particular.

LoRa’s CCS modulation encodes each symbol in spreading factor (SF) bits with a unique chirp frequency trajectory. With chirp bandwidth fixed (BW = 125, 250 or 500 kHz), LoRa creates virtual channels by varying the chirp duration based on changing the SF bits, (i.e., $T_c = 2^\textrm{SF}/\textrm{BW}$), while supporting SFs of 7 to 12. A higher SF means lower data rate, however, better robustness against noise i.e., better coverage~\cite{Aamir}.  

On the other hand, LoRaWAN defines open specifications of a complete networking solution based on LoRa radio technology.  {Like cellular networks, a LoRaWAN network topology is a hybrid of wireless and wired star-of-stars, consisting of multiple gateways to tunnel into (from) a wired backhaul of uplink (downlink) messages of (to) end devices (EDs). LoRaWAN specifications add constraints on the LoRa physical layer parameters according to regional specifications. For instance, in Europe, the permissible bandwidth is $\textrm{BW}\in\{125,250\}$ kHz and $\textrm{SF}\in[7...12]$, therefore, the actual data rates range from about 300 bps to 11 kbps.}

\subsection{LoRaWAN for Positioning}
What makes a LoRaWAN specifications unique for localization use-cases is:
\begin{itemize}[leftmargin=*]
\item \textbf{Rx Macro-diversity:} Receive macro-diversity is a unique feature of LoRaWAN networks, in which any physically apart gateways can receive uplink messages. The LoRa EDs transmit a message in an uplink broadcast mode that is not specific to a gateway. Any gateway receiving the message, forwards it to the server, which takes care of removing duplicates and selects the gateway for downlink response (if any) based on the best RSSI estimates. Apart from improving coverage and capacity, macro-diversity is an attractive feature for building  TDoA or RSSI-based localization in LoRaWANs. 
\item \textbf{Roaming}: With standardized roaming architecture by LoRa Alliance, passive roaming allows devices to use other network’s gateways and forward messages between different LoRaWAN operators. Thus, roaming will improve coverage as well as enhance localization feasibility in LoRaWAN networks.
\end{itemize}

This simplified, flexible and standardized network architecture, and robust LoRa PHY, makes LoRaWAN networks extremely attractive for geolocation: localization, tracking and geo-fencing without GPS, which reduces both the cost and the energy consumption. Besides, LoRaWAN is also suitable for localization in indoor environments~\cite{c13}.

\begin{figure*}[!t]
\centering
\includegraphics[width=1\linewidth]{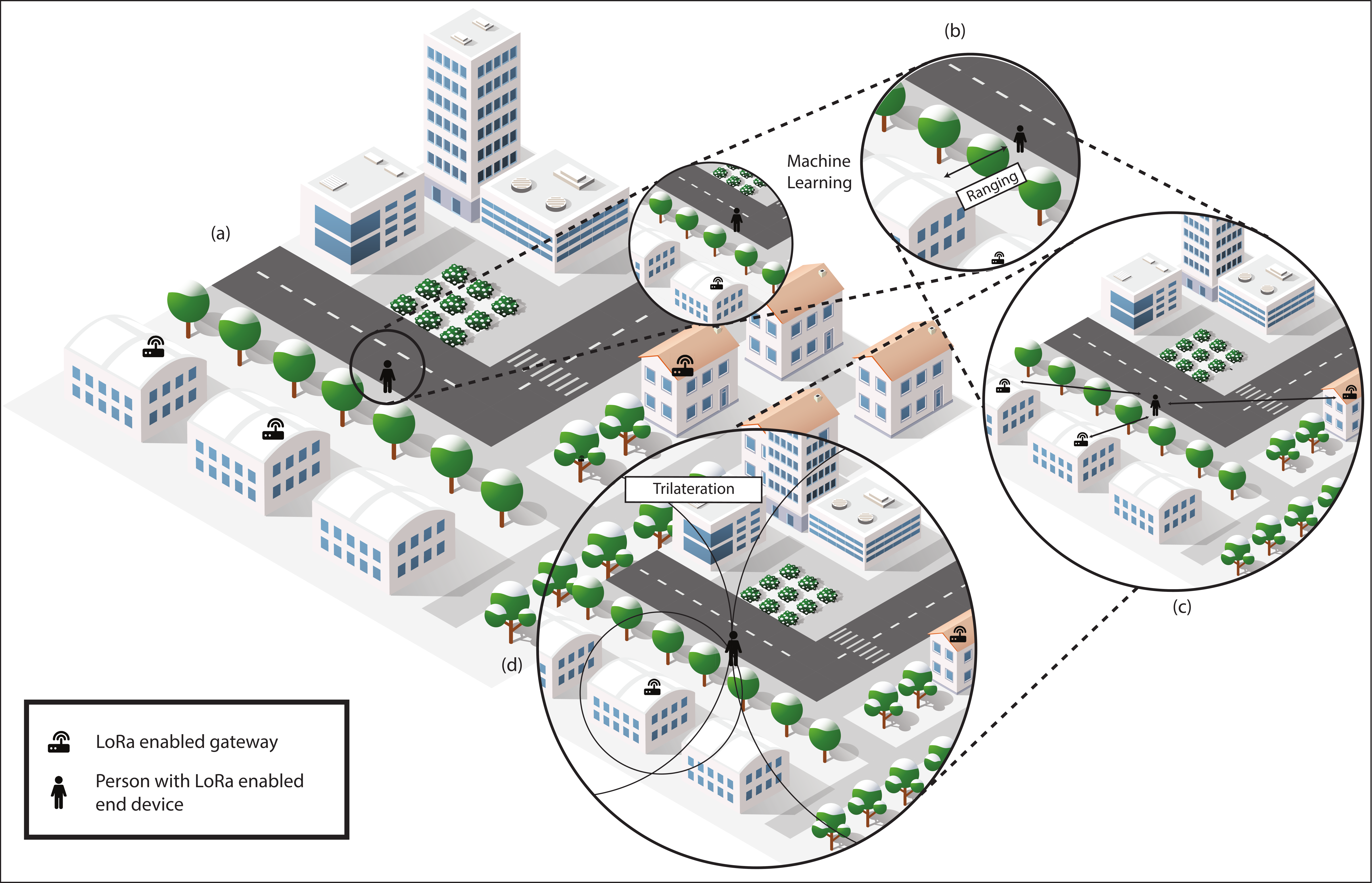}
\caption{A sketch of LoRa position system: (a) Signal reception from end-device, (b) Ranging on one gateway, (c) Signal reception and ranging on three gateways, (d) Trilateration}
\label{fig5}
\vspace{-10pt}
\end{figure*}

\section{LoRa Positioning System}
 {The reference points of a positioning system essentially emulate the satellite navigation system by replacing satellites with BSs. Similarly, we can construct an LPS in which LoRa gateways act as pseudo-satellites to track LoRa enabled end-devices} (see Fig.~\ref{fig5}).  {In particular, the LPS system works as follows.  
\begin{enumerate}
    \item LoRa gateways measure the signal strength of the received messages from end-devices.
    \item The gateways calculate their respective Euclidean distances based on RSSI-to-distance relationship.
    \item Using the distance estimate from $(N+1)$ gateways, LPS performs trilateration to find the coordinates of a device in $N$-dimensional space, while this study considers three gateways only.
\end{enumerate}
}
These steps are illustrated in steps (a)-(d) in Fig.~\ref{fig5}. Two vital aspects of the LPS system are data collection setup and RSSI-to-distance mapping methodology. We describe each in the rest of this section.

\subsection{Data Collection}
Since RSSI characteristic is propagation environment-dependent, a dataset of RSSI for known locations is required to build or train RSSI-to-distance mapping models. The models are then used to perform the inverse function of translating measured RSSI to distances.
\subsubsection{Measurement setup}
Our setup consists of three open-source single channel LoRa gateways (Dragino LG01-S) and a LoRa end-device (Dragino LoRa shield). It operates at 433 MHz on SF 12, as it offers the largest area of coverage. The end-device is programmed to broadcast a character for a minute. It is then halted and a different character is broadcasted for another minute. This process is used to collect RSSI data at different locations. The gateways are installed in a triangular manner such that each gateway experiences different propagation conditions to the ED. Each gateway is connected to a processing device, which logs the RSSI values and the corresponding character. The locations of the data collection points are marked digitally on a bird's-eye view of the data collection site.

\subsubsection{Dataset structure}
The raw dataset of each gateway has one entry for each received message, containing two values: an identifier and the RSSI. The distance between the data collection point and the reference gateway is calculated using a digitally marked map. The final dataset contains 2 columns per gateway; the RSSI and the corresponding distance. It is important to remark that the dataset is noisy and unbalanced, i.e., the values and number of RSSI entries for each point are different. Later, the dataset is fitted using different models to develop RSSI-to-distance functions for each gateway.

\subsection{RSSI-to-Distance Mapping }
This sub-section develops the necessary background on considered ranging models, including: the standard path loss model, the traditional regression models, and modern machine learning models. 

The standard path loss model is often represented by a log-distance power law in which the path loss is logarithmically related to the distance, and the power of distance is termed as the propagation exponent. The best fit of the collected data points with the path loss model requires finding the propagation exponent that minimizes the root mean square deviation of the data points with the model. However, the standard path loss model is severely affected by the propagation conditions, and the range of operation. A highly variable ranging function can not be estimated using the traditional path loss models. Thus, we studied various regression and machine learning models to accommodate the variability of the ranging function. 

\subsubsection{Regression Models}  
To find a regression-based ranging function, we used linear, polynomial, exponential, and Gaussian models (see \cite{ML_gen}).  {\textit{Linear regression} is a low-complexity and fast technique to fit a line to the training dataset. It is attractive since the RSSI-to-distance mapping, on inspection, revealed a linear trend for smaller transmit receive (TR) distances. Whereas, \textit{polynomial regression} fits a model of degree $\geq 2$ on the dataset with simple empirical relations. Polynomials are unbounded, oscillatory functions that are sensitive to noise. In this study, we used polynomial models of degree three.}

\textit{Exponential regression} finds the exponential function that fits best with the dataset. The exponential relation naturally arises from the power law of the path loss model.  {We used exponential models containing two coefficients. Note that, the uniform nature of the exponential function proves inefficient for dataset that include multiple points of inflection and, consequently, multiple concavities.} Finally, the \textit{Gaussian regression}, assuming the dataset is sampled from a normal distribution, uses lazy learning and a kernel function to predict the value of unseen test points.  {We employed the Gaussian model with multiple exponential terms, and tuned their amplitudes, centroids and peak widths to fit the dataset. 

\subsubsection{Machine Learning (ML) Models} We used non-parametric ML models to find a suitable ranging function, which included smoothing spline, decision trees, random forest, and support vector regression (SVR) (see \cite{ML_gen}).  {\textit{Smoothing spline models} are ideal for a noisy dataset. A $k$-th order spline is a piecewise polynomial function of degree $k$. The smoothing spline functions perform regularized regression, using inputs as knots and minimizing overfitting by damping the coefficients of the estimation function. In this study, we used cubic splines as the piecewise polynomial blocks with high smoothing parameters to accommodate the noise in the training dataset.} 

The \textit{decision trees} and \textit{random forest} algorithms are decision support tools. They use tree-like models of decisions and consequences, while their nested if-else structure makes them ideal for the unbalanced and noisy dataset.  {We utilized boosted trees to model the random forest algorithm. Boosted trees are combined along with successive decision trees, and the results are compiled along the way. We use 30 trained learners and tree pruning in our boosted trees.}

\textit{SVR} tries to fit the error within a certain threshold. The best fit is the hyperplane that has the maximum number of points in the threshold.  {These models use a linear kernel function with sequential minimal optimization.} SVRs are ideal for complex relations and have long training times.

\begin{table*}[!t]
\renewcommand{\arraystretch}{1.2}
\caption{Training Accuracy of different positioning models}
\label{acc}
\centering
\begin{tabular}{|c|c|l|c|c|}
\hline
\textbf{Location} & \textbf{Type} & \textbf{Model} & \textbf{Error (m)} & \textbf{Accuracy (\%)}\\
\hline
\multirow{9}{*}{Outdoor} &\multirow{5}{*}{Regression models}  &Path loss model&45.78&77.21\\ 
\cline{3-5}
& &Linear regression&45.75&77.24\\
\cline{3-5}
& &Polynomial&46.68&76.30\\
\cline{3-5}
& &Exponential&45.78&77.21\\
\cline{3-5}
& &Gaussian&46.02&76.97\\
\cline{2-5}
&\multirow{4}{*}{Machine learning}  
&Smoothing spline&36.29&85.68\\
\cline{3-5}
& &Support vector regression&40.44&82.21\\
\cline{3-5}
& &Decision trees&42.55&80.31\\
\cline{3-5}
& &Random forest&46.71&76.27\\
\hline
\multirow{9}{*}{Indoor} 
&\multirow{5}{*}{Regression models}
&Path loss model&14.12&81.69\\
\cline{3-5}
& &Linear regression&13.75&82.64\\
\cline{3-5}
& &Polynomial regression&12.72&85.14\\
\cline{3-5}
& &Exponential&14.12&81.69\\
\cline{3-5}
& &Gaussian&12.75&85.07\\
\cline{2-5}
&\multirow{4}{*}{Machine learning}  
&Smoothing spline&09.38&91.92\\
\cline{3-5}
& &Support vector regression&13.75&82.64\\
\cline{3-5}
& &Decision trees&11.45&87.96\\
\cline{3-5}
& &Random forest&12.53&85.58\\
\hline
\end{tabular}
\end{table*}

 {ML and regression models use minimization algorithms to decrease the error of the estimated fit on the dataset. Minimization algorithms determine the conditions for the lowest value of a given function. We use the trust-region, sequential minimal optimization (SMO), and least square boosting (LBoots) algorithms to develop the ranging models. The \textit{trust region} algorithm (used in exponential, Gaussian, and path loss models) considers a simplified approximation of the objective function and minimizes it over a sub-region. If the minimization does not increase accuracy, then the sub-region is contracted. \textit{SMO}, which we used in SVR, is designed to solve the quadratic programming problem, essential for the training of SVRs. SMO breaks the quadratic function into simpler problems and then solves them analytically to obtain the solution~\cite{ML}. Finally, in boosted trees, we used \textit{LSBoost} that reduces the mean square error of the model in ensemble learning techniques. It works by fitting an ensemble to minimize the difference between the actual value of the prediction made by the aggregation of all the previously joined ensembles.}

\subsection{Trilateration }
The distances provided by the ranging models are fed into the trilateration algorithm along with the coordinates of the pseudo-satellites. Trilateration extends circles from each of the reference locations with radii equal to the estimated distances from the unknown point. These circles intersect where the unknown point is located. Because of the  errors in the estimation, the circles might not intersect at the same point, but in this case, an area is isolated by the intersections where the unknown point can be found. In the following section, we discuss the assembly and implementation of our positioning system in two different environments.

\section{Case Studies:Results and Discussion}
\label{sec:CaseStudy}
We studied the performance of earlier described ranging models in estimating the distance of a node and, eventually, its position. For this purpose, we deployed LPS in outdoor and indoor environments, first to collect RSSI data and train the ranging models and then to validate the models in finding the unknown location of a transmitting device. Our data collection and testing methodology is as  follows.

\subsection{Outdoor Positioning }

We installed gateways at different outdoor locations, covering an area of approximately 25,753 m$^{2}$ in which we collected dataset at 26 locations. The gateways and the dataset collection points are marked in Fig. \ref{fig_sim4}. In this plan, with reference to the data collection site, gateway A was farthest, gateway B was most obstructed, and gateway C had the clearest view. We collected a dataset of 4181 tuples, and fed it to the ranging models for fingerprinting. As compared to others, the smoothing spline model provided the minimum ranging error at any gateway, with the root mean squared error (RMSE) of:
\begin{itemize}
\item Gateway A: 46.60~m,
\item Gateway B: 49.95~m,
\item Gateway C: 27.74~m.
\end{itemize}

We tested the positioning accuracy of these ranging models by measuring the Euclidean distance between true coordinates and predicted coordinates of a test point. Table~\ref{acc} shows the positioning accuracy for different ranging models, where smoothing spline has the highest mean absolute accuracy of 36.29~m and mean percentage accuracy of 87.24\%.

  {For all the ranging models in Table~\ref{acc}, gateway C has the least while gateway B has the most training error. The range of distances for the training data of outdoor positioning are;
\begin{itemize}
\item Gateway A: 79.11~m – 284.43~m
\item Gateway B: 21.97~m – 210.56~m
\item Gateway C: 16.76~m – 203.06~m
\end{itemize}
Additionally, the mean RSSI of the dataset for gateway C is better than that of gateways A and B. The mean RSSIs are: $\{\textrm{A}{:} -97.32, \textrm{B}{:} -92.75, \textrm{C}{:} -89.92\}$ dBm. It shows that the training dataset for gateways A and B cover more geographical area than gateway C, and it can be inferred that the closer the area-of-interest is to the gateways, the better the results of training-based ranging models.}

 {To test ranging/positioning performance of trained models, we selected five tests points (TP1--TP5) in the data collection site. Test points are selected such that TP1 and TP2 have more distance from gateway A than TP3, TP4, and TP5 (see Fig.~\ref{fig_sim4}). Based on the average RSSI of a test point, the ranging error contributions of gateways for different test points are shown in Fig. \ref{fig_sim2}. It is observed that, for TP3, TP4, and TP5, the error contributions from gateway A are comparable. Error in TP1 is higher than the rest, and TP2 is elevated to be in LOS with A, leading to the least error contribution from A. On the other hand, gateway B has comparable errors for all points except TP2, which shows a higher error due to the tree cover with B. Gateway C has a NLOS path to TP2, which reflects in the ranging error.}

\begin{figure}[!t]
\centering
\includegraphics[width=1\linewidth]{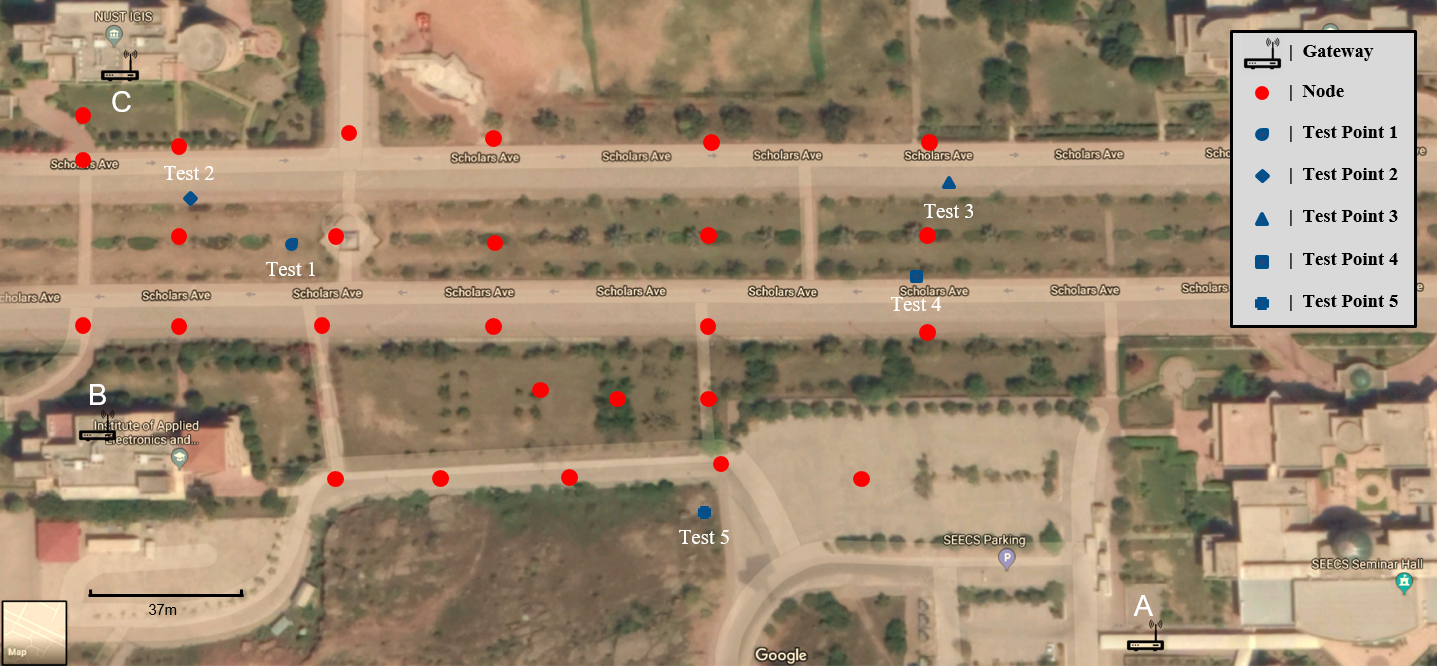}
\caption{A bird's eye view of outdoor data collection points}
\label{fig_sim4}
\end{figure}

\begin{figure}[!t]
\centering
\includegraphics[width=1\linewidth, height=0.8\linewidth]{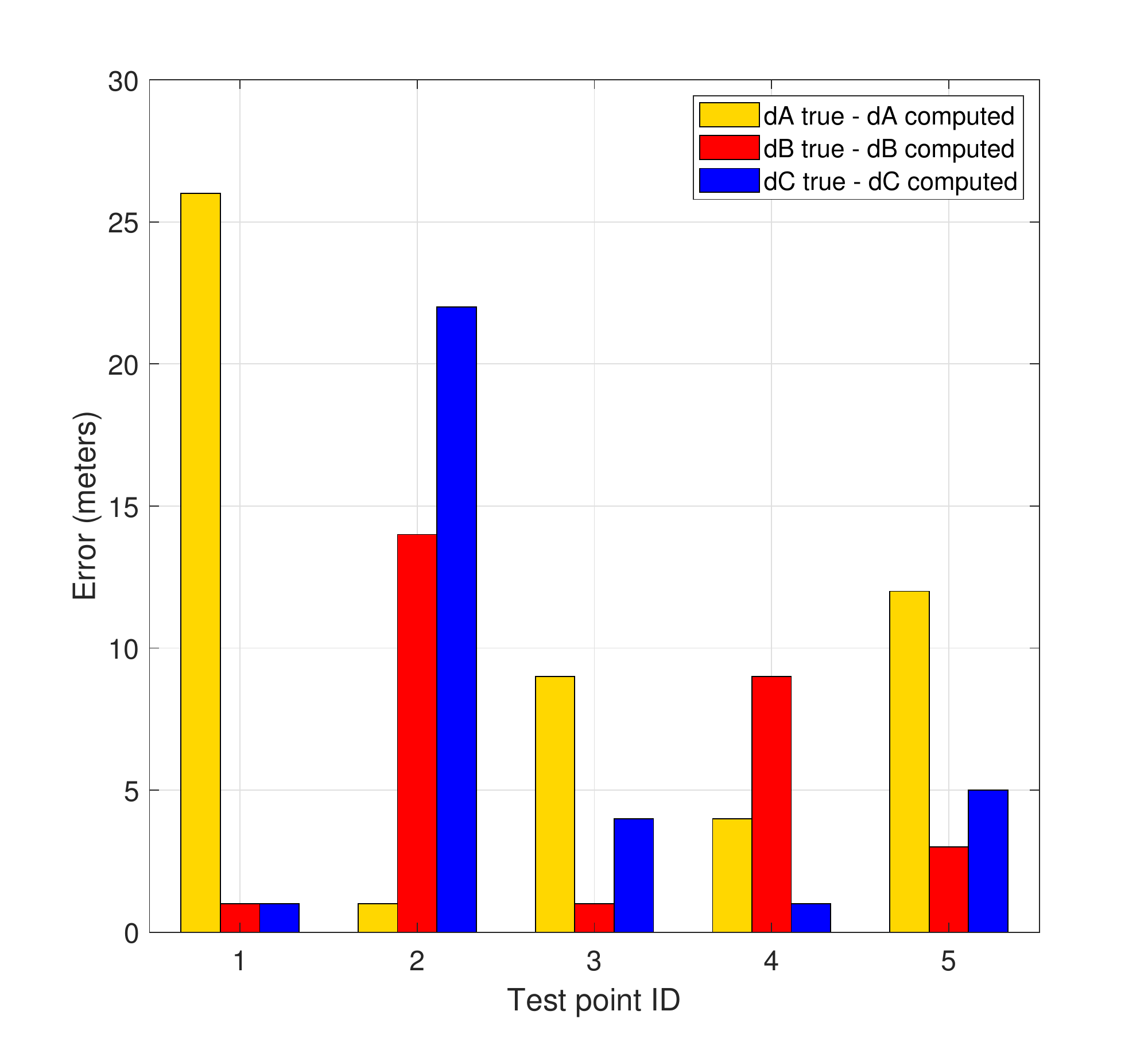}
\caption{Ranging error between the true and calculated distance of test points from the gateways for outdoor positioning}
\label{fig_sim2}
\end{figure}

\subsection{Indoor Positioning }
To evaluate indoor ranging and positioning accuracy of LPS, we deployed gateways that bounded the data collection site of 25 points within an area of  approximately 3,422 m$^{2}$ (see Fig. \ref{fig_sim3}). The gateways’ deployment was such that: gateway D experienced the tunneling effect, gateway F was on the first floor, and gateway E had more obstacles to the data collection points. The ranging models were trained using a dataset of 968 tuples collected in this setup. Again, the smoothing splines provided the best RSSI-to-distance mapping, with RMSE:
\begin{itemize}
\item Gateway D: 10.03 m,
\item Gateway E: 14.24 m,
\item Gateway F: 11.75 m.
\end{itemize}
Using the same positioning methodology as in outdoor setup, the accuracy of the different indoor positioning models is given in Table~\ref{acc}, with smoothing splines giving the highest mean absolute accuracy of 9.38~m and mean percentage accuracy of 84.60\%.

 {All ranging models provided the least training error for gateway D. The range of distances for indoor positioning are:
\begin{itemize}
\item Gateway D: 3.54~m – 48.83~m
\item Gateway E: 4.58~m – 60.91~m
\item Gateway F: 5.33~m – 55.30~m
\end{itemize}
The mean RSSI at the gateways are: $\{\textrm{D}{:} -55.32, \textrm{E}{:} -66.82, \textrm{F}{:} -56.93\}$ dBm. Therefore, our earlier inference that the training-based models perform better for a smaller range of distances also holds for the indoor scenario, which is also reflected in the testing phase.} 

\begin{figure}[!t]
\centering
\includegraphics[width=1\linewidth]{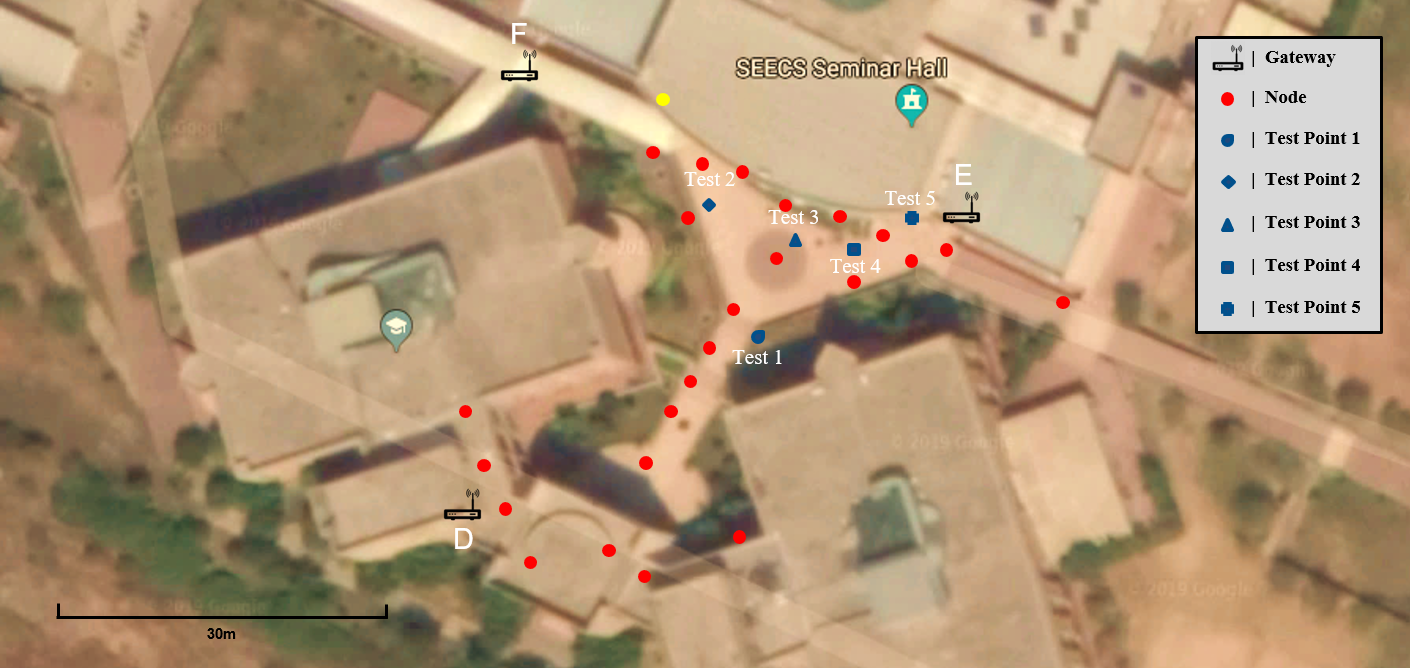}
\caption{A bird's eye view of indoor data collection points}
\label{fig_sim3}
\end{figure}
\begin{figure}[!t]
\centering
\includegraphics[width=1\linewidth, height=0.8\linewidth]{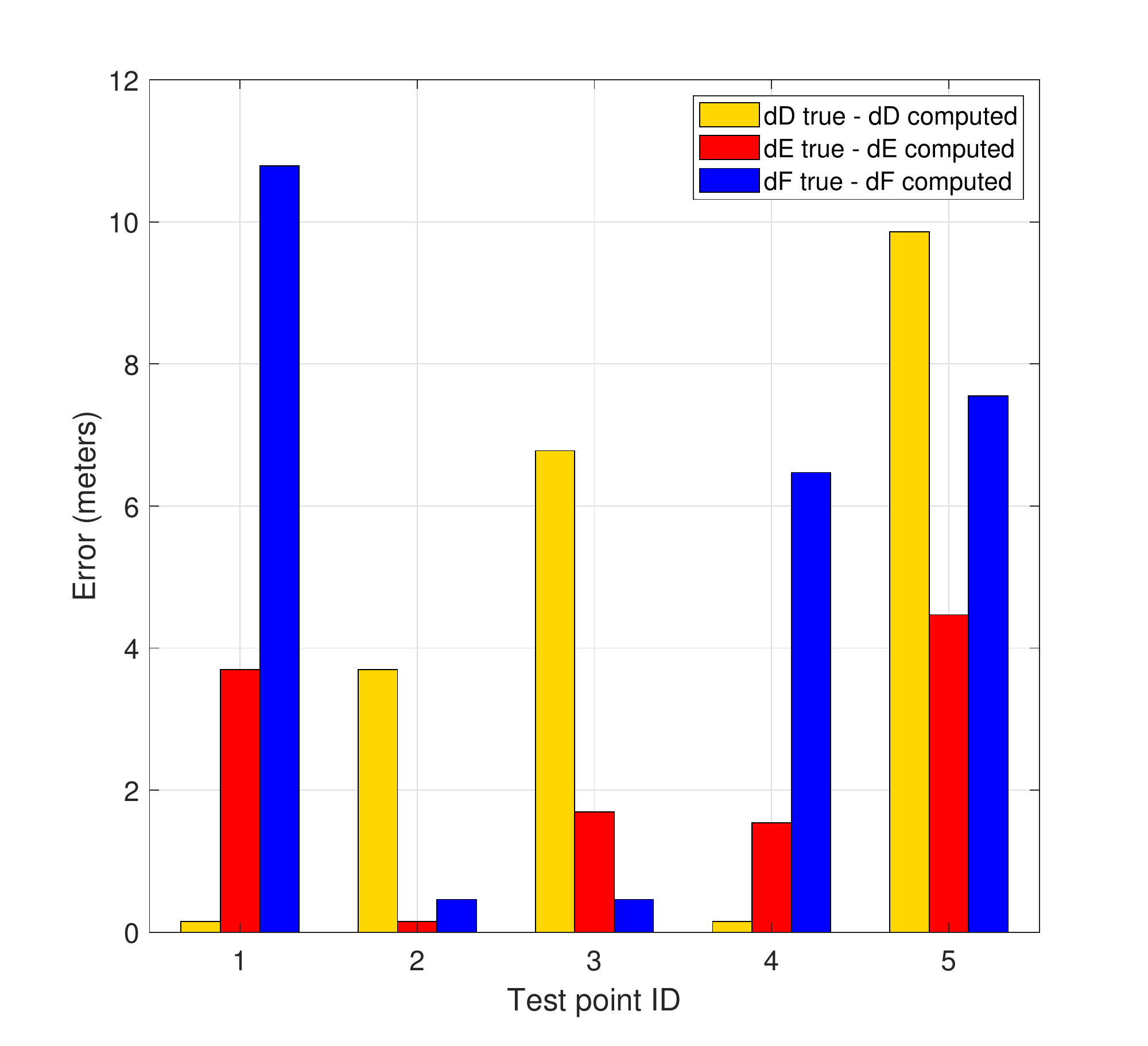}
\caption{Ranging error between the true and calculated distance of test points from the gateways for indoor positioning}
\label{fig_sim}
\end{figure}

 {The selected testing points are marked in Fig. \ref{fig_sim3}, while their respective ranging errors are shown in  Fig. \ref{fig_sim}. Note that TP1 is in direct LOS of gateway D, and thus, has the least error contributed from D as compared to TP2, TP3, and TP5, which experience significant error from D for being in NLOS. The error contribution of D for TP5 is higher due to the larger distance. In general, TP5 is surrounded by obstacles, and has significant error from all gateways. TP1--TP4 are in a similar NLOS path from E, and thus show comparable error contributions. Gateway F has multiple obstacles between itself and TP1, TP4 and TP5, resulting in significant error contributions from F. TP1 has a wide concrete boundary in its path towards F, which increases the error in distance computation. The TR distance between F and TP4 and TP5 is much higher than TP2 and TP3, which reflects in the results.
}

\subsection{Discussion}
 {In \cite{c10},  trilateration is performed using RSSI fingerprinting of BLE beacons from three closest anchor nodes. The authors reported that trilateration returned a range of values instead of a single value, which is also observed in our system. The authors achieved an accuracy of 5.06~m within an area of 290.4~m$^2$. The relative precision of their system is 27.69\%; i.e. the model accurately indicates an area of 27.69~m$^2$ for any unknown point, in an area of 100~m$^2$. Our method achieves an accuracy of 9.38~m in an area of 4011.36~m$^2$ (65.76~m x 61.0~m). The relative precision of our system is 6.8\%. It may be noted that the range of Bluetooth is less than LoRa, therefore, the percentage accuracy is better for LoRa technology.} 

 {In \cite{c6}, the authors used Wi-Fi based localization to estimate the position of a moving device using a topological radio-map. A hidden Markov model was applied to fuzzy decision trees to obtain an accuracy of 33.8~m for multi-floor, and 2.5~m for same-floor localization, using 126 access points. Our model provides an accuracy of 9.38~m using three gateways.}

  {In \cite{c7}, the authors used ZigBee for RSSI-based fingerprinting and room-level localization. This approach involved the use of devices in 40~m$^2$ rooms, to localize users. The authors reported an accuracy of 85\%-98.2\% depending on the number of receivers, that varies from 10 to 25. In contrast, we achieve an accuracy of 93.1\% with three receivers.}

 {In~\cite{c5}, Sigfox is used to explore class-based and intra-class localization, where former classifies a group of nodes and the latter classifies nodes within a group. For class-based localization,  the authors achieved accuracy ranging from 78\%-100\%. They observed an accuracy of 85\% in the intra-class setup. Our outdoor models provide a percentage error of 12.75\% on all accounts. The mean absolute accuracy is 36.29~m in an area of 28,892~m$^2$ (233~m x 124~m), which gives a mean percentage outdoor accuracy of 85.68\%. The difference in percentage accuracy of outdoor and indoor environments is compliant with the path loss model, which states a decrease in accuracy with an increase in distance.}

\section{Conclusion and Future Directions}
This article explored the use of regression and machine learning (ML) models for RSSI-fingerprinting in LoRa networks. Based on the amount of data, the results showed that the ML approach towards RSSI fingerprinting provides promising results. We provided two case studies for the trilateration-based LPS. The indoor implementation provided comparable results to the outdoor positioning system. This makes LoRa a feasible solution to the limitations of satnav systems.  Because of  the train-then-test methodology, different ranging models can be trained for different environmental conditions, thus creating a more robust positioning system. 

Environmental factors, weather conditions and obstacles such as walls and buildings greatly influence signal propagation and quality. This dependence on the environment-of-operation can be accommodated by training different models for each gateway, in different environmental conditions. Asset tracking, medical supplies and patient tracking, activity monitoring in old homes, child protection and vehicle tracking can benefit greatly from the low-power and long-range nature of LoRa networks.}

The performance is expected to improve further by using deep learning models. A larger dataset is expected to increase the precision of the LPS. A high-precision GNSS ensemble can be used to improve the training dataset. High gain antennas would not only improve the range of LPS, but provide better RSSI variability, which, consequently, will provide better results. The overall accuracy greatly depends upon the positioning algorithm. It can be improved by using multilateration using an assembly of multiple reference gateways.  Numerous ML approaches can be adopted for the localization of the unknown points. Additionally, many safety and security systems can be designed, by taking advantage of localization. These developments will require extensive research, experimentation, and testing before they can be used in practical implementations. The recent advances in the field of ML can enable us to design practical systems that provide great benefit to the IoT ecosystem.

\bibliographystyle{ieeetran}

\end{document}